\documentclass[twocolumn,showpacs,superscriptaddress,preprintnumbers,prd]{revtex4}
\usepackage{graphicx,amsmath,dcolumn,bm}
\begin{document}
\preprint{KEK-TH-1208}
\title{\Large \bf Determination of gluon polarization \\
                  from deep inelastic scattering and collider data}
\author{M. Hirai}
\affiliation{Department of Physics,
             Tokyo Institute of Technology,
             Meguro, Tokyo, 152-8551, Japan}
\affiliation{Department of Physics, Juntendo University, 
             Inba, Chiba, 270-1695, Japan}
\affiliation{Department of Physics, 
             Tokyo University of Science, Noda, Chiba 278-8510, Japan}
\author{S. Kumano}
\affiliation{Institute of Particle and Nuclear Studies,
          High Energy Accelerator Research Organization (KEK) \\
          1-1, Ooho, Tsukuba, Ibaraki, 305-0801, Japan}
\affiliation{Department of Particle and Nuclear Studies,
           Graduate University for Advanced Studies \\
           1-1, Ooho, Tsukuba, Ibaraki, 305-0801, Japan}     
\date{November 27, 2008}
\begin{abstract}
We investigate impact of $\pi^0$-production data at Relativistic Heavy Ion
Collider (RHIC) and future E07-011 experiment for the structure
function $g_1$ of the deuteron at the Thomas Jefferson National
Accelerator Facility (JLab) on studies of nucleonic
spin structure, especially on the polarized gluon distribution function.
By global analyses of polarized lepton-nucleon scattering and
the $\pi^0$-production data, polarized parton distribution functions
are determined and their uncertainties are estimated by the Hessian method.
Two types of the gluon distribution function are investigated.
One is a positive distribution and the other is a node-type distribution
which changes sign at $x \sim 0.1$. Although the RHIC $\pi^0$ data seem to
favor the node type for $\Delta g(x)$, it is difficult to determine
a precise functional form from the current data. However, it is interesting
to find that the gluon distribution $\Delta g(x)$ is 
positive at large $x$ ($>0.2$) due to constraints from
the scaling violation in $g_1$ and RHIC $\pi^0$ data.
The JLab-E07-011 measurements for $g_1^d$ should be also able to reduce
the gluon uncertainty, and the reduction is comparable to the one
by RUN-5 $\pi^0$-production data at RHIC.
The reduction is caused mainly by the error correlation between polarized
antiquark and gluon distributions and by a next-to-leading-order (NLO)
gluonic effect in the structure function $g_1^d$.
We find that the JLab-E07-011 data are accurate enough to probe
the NLO gluonic term in $g_1$.
Both RHIC and JLab data contribute to better
determination of the polarized gluon distribution in addition to
improvement on polarized quark and antiquark distributions. 
\end{abstract}
\maketitle

\section{Introduction}
\label{intro}
Quark and gluon contributions to the nucleon spin are described by 
polarized parton distribution functions (polarized PDFs)
and their first moments. It became clear that only a small fraction
of nucleon spin is carried by quarks and antiquarks. Therefore,
a large gluon polarization or effects of orbital angular momenta
should be possible sources for explaining the origin
of the nucleon spin. The gluon polarization is expected to be clarified
in the near future, whereas it would take time to determine the effects
of the orbital angular momenta.

Polarized PDFs have been investigated by global analyses of data on
polarized lepton-nucleon deep inelastic scattering (DIS) and
proton-proton collisions \cite{ppdfs-1,bb02,lss06,aac,aac06,dssv08}.
Polarized quark distributions are determined relatively well;
however, the polarized gluon distribution $\Delta g(x)$ is not 
accurately determined. 
Here, $x$ is the Bjorken scaling variable, and $\Delta g (x)$ 
is the difference between the gluon distribution with
helicity parallel to that of parent nucleon and the one
with helicity anti-parallel.

The gluon distribution contributes to the structure function $g_1$
as a higher-order effect in the expansion by the running coupling
constant $\alpha_s$ of quantum chromodynamics.
The unpolarized gluon distribution has been determined primarily by
the $Q^2$ dependence of $F_2$ at small $x$, where $Q^2$ is defined 
by the momentum transfer $q$ by $Q^2=-q^2$ in lepton scattering.
The kinematical range of $x$ and $Q^2$ is still limited for $g_1$
in determining $\Delta g(x)$ by the scaling violation, so that
the determination of $\Delta g(x)$ is difficult from the scaling violation.
Nonetheless, it is noteworthy that there are $Q^2$ differences between
the COMPASSS and HERMES data for $g_1$ in the range of $x\sim 0.05$.
Such $Q^2$ differences could be used for constraining a gluon
polarization at large $x$ as pointed out in Ref. \cite{aac06}.
However, this idea should be tested by future measurements on
the scaling violation in $g_1$ because the $Q^2$ differences
could originate also from higher-twist effects \cite{lss06}.

Other types of measurements are needed to improve the situation 
of $\Delta g(x)$. There were measurements on $\Delta g(x)$
in lepton-nucleon scattering by observing high-$p_T$
hadrons \cite{hermes-dg-00-07,smc-dg-04,compass-dg-06} and
open-charm events \cite{compass-dg-08}. These data provided
constraints on the gluon polarization at $x \sim 0.1$.
They indicated that the ratio $\Delta g(x)/g(x)$ is small
although experimental errors are still large. Measurements
at Relativistic Heavy Ion Collider (RHIC) are also important
for constraining the gluon polarization. For example,
$\pi^0$ and jet-production data \cite{phenix-pi,star-jet}
in polarized proton-proton collisions are valuable
for the determination of $\Delta g(x)$.
In fact, we showed that the $\pi^0$ data play an important
role in reducing the uncertainty of $\Delta g(x)$ by a global
analysis including the $\pi^0$ data in addition to the DIS data
\cite{aac06}. Certain fragmentation functions are used
in the analysis of $\pi^0$-production processes. We should be
careful that gluon and light-quark fragmentation functions have
large uncertainties at small $Q^2$ or small $p_T$ \cite{hkns07}.

In order to determine the polarized PDFs including the gluon polarization,
precise measurements are needed also in DIS.
After our previous analysis \cite{aac06},
new DIS data are reported by the CLAS \cite{clas-06}, HERMES \cite{hermes-07},
and COMPASS \cite{compass-07} collaborations. In future, the structure
function $g_1$ for the deuteron will be accurately measured at 
the Thomas Jefferson National Accelerator Facility (JLab)
by the proposed experiment E07-011 \cite{jlab-pr-07-011, JLabD-ProData}. 
The measurements at JLab should be valuable for
reducing large uncertainties of the polarized antiquark and
gluon distributions because they cover a large-$x$ region at small $Q^2$.

In this work, we determine the polarized PDFs by global analyses
of the data for spin asymmetries $A_1$ in polarized lepton-nucleon DIS
and double spin asymmetries $A_{LL}$ of the $\pi^0$ production
in proton-proton collisions. 
In particular, we focus our discussions on determination
of the polarized gluon distribution function.
The purposes of this article are the following.
\begin{itemize}
\item[(1)]
The RHIC $\pi^0$-production data could indicate a node-type distribution
for $\Delta g(x)$, so that two types of distributions are studied
in the global analysis by taking a positive distribution for $\Delta g(x)$
and a node-type distribution which changes sign at $x \sim 0.1$.
\item[(2)]
Among various future experimental projects, we investigate impact 
on the polarized PDFs, especially on the polarized gluon distribution
function, from precise measurements on the structure function $g_1$
of the deuteron by the JLab-E07-011 experiment
\cite{jlab-pr-07-011, JLabD-ProData}.
The E07-011 data will be so precise that $\Delta g(x)$ could be constrained. 
\item[(3)]
The polarized PDFs obtained with the E07-011 data are compared
with the ones including the $\pi^0$-production data especially
in reducing uncertainties of $\Delta g(x)$ and its first moment. 
\end{itemize}
We provide a new AAC08 (Asymmetry Analysis Collaboration in 2008) 
library by the current global analyses because updated AAC distributions
have not been provided since the 2003 version \cite{aac,aac-web}. 

This article is organized as follows. In Sec. \ref{analysis}, three data sets
are introduced for investigating roles of the RHIC $\pi^0$ and JLab data,
and our analysis method is explained.
Analysis results are discussed in Sec. \ref{results},
and they are summarized in Sec. \ref{summary}.

\section{Analysis method}
\label{analysis}

\begin{table}[t]
\caption{Used data sets in our global analyses.
         The ``DIS" indicates current DIS data on $g_1$. 
         The RHIC $\pi^0$ data are taken from Ref. \cite{phenix-pi}.
         The JLab-E07-011 data are calculated by using estimated errors in 
         Refs. \cite{jlab-pr-07-011, JLabD-ProData} and set-A results,
         and actual vales are shown in Table \ref{Tabl:JLab-E07-011}.
         The notation $\bigcirc$ indicates an included data set.}
\label{Tabl:JLabD}
\begin{center}
\begin{tabular}{cccc} 
\hline
\ Analysis set \ 
        &  \ DIS \       & \ RHIC $\pi^0$   & \ E07-011 \     \\ \hline
A       &  $\bigcirc$    & $-$              & $-$             \\
B       &  $\bigcirc$    & $\bigcirc$       & $-$             \\
C       &  $\bigcirc$    & $-$              & $\bigcirc$      \\ \hline
\end{tabular}
\end{center}
\end{table}
In order to investigate roles of the RHIC $\pi^0$ and E07-011 data,
three data sets in Table \ref{Tabl:JLabD} are prepared.
The notation ``DIS" indicates all the present DIS data on $g_1$,
and they are almost the same data used in the AAC06 analysis \cite{aac06}.
There are minor changes. The same data are used for
SLAC, EMC, SMC, and HERMES; however, updated ones are used
for the deuteron by the COMPASS \cite{compass-07}.
In addition, the CLAS data \cite{clas-06} are included in the DIS set.
This is called set A; the one with $\pi^0$ data is set B;
the one with the E07-011 is set C.
The data for the RHIC $\pi^0$ and expected E07-011 are taken
from Refs. \cite{phenix-pi} and \cite{jlab-pr-07-011, JLabD-ProData},
respectively. The RUN-5 data are used for the $\pi^0$ production.
Preliminary RUN-6 data by the PHENIX collaboration \cite{phenix-run6}
and jet data of the STAR collaboration \cite{star-jet} 
are not included in our analyses. We may include these data in our
future global analysis.
Comparing three analysis results, we try to clarify roles of
the RHIC $\pi^0$ and E07-011 data in the determination of the polarized PDFs,
especially on $\Delta g(x)$, in comparison with the results by the set A.

The expected data for $A_1^d$ of the E07-011 experiment are estimated
in the following way. The asymmetries are assumed to be the same as
the ones of the analysis A: 
$A_1^d (x,Q^2)_{\text{E07-011}} \equiv A_1^d (x,Q^2)_{\text{set-A}}$.
Expected errors of the E07-011 experiment are estimated 
for $(\delta g_1^d /g_1^d)_{\text{E07-011}}$ in the proposal  
\cite{jlab-pr-07-011, JLabD-ProData}, and they are converted
to the errors of $A_1^d$ by
\begin{equation}
  \delta A_1^d (x, Q^2) _{\text{E07-011}}
  \equiv \bigg [ \frac{\delta g_1^d(x,Q^2)}{g_1^d(x,Q^2)} \bigg ]_{\text{E07-011}} 
  A_1^d(x,Q^2)_{\text{set-A}}.
\end{equation}
Estimated values and their errors are shown for the asymmetry $A_1^d (x, Q^2)$
in Table \ref{Tabl:JLab-E07-011}.
It should be noted that there are two solutions for the polarized PDFs,
the positive and node-type distributions, so that there are two
sets of $A_1^d$ data in Table \ref{Tabl:JLab-E07-011}.

\begin{table}[t]
\caption{Expected asymmetries $A_1^d (x,Q^2)$ and their errors in
the E07-011 experiment. They are calculated by using
the set-A analysis results and estimated errors
$\delta g_1^d /g_1^d$. The details are explained in the text.}
\label{Tabl:JLab-E07-011}
\begin{center}
\begin{tabular}{ccccccc}
\hline
$x$ & $Q^2$      & $\delta g_1^d/g_1^d$ & $A_1^d$ & $\delta A_1^d$
                                             & $A_1^d$ & $\delta A_1^d$ \\ 
    & (GeV$^2$)  &                      &                                      
                       \multicolumn{2}{c}{positive} & \multicolumn{2}{c}{node} 
\\ \hline
0.175 \  & \ \ 1.4 \ \ & \ \ 0.020  \ \ 
                         & \ \ 0.0841 \ & \ 0.0017 \ \
                         & \ \ 0.0819 \ & \ 0.0016 \\
0.25 \ \ & \ \ 1.9 \ \ & \ \ 0.014  \ \
                         & \ \ 0.1568 \ & \ 0.0022 \ \
                         & \ \ 0.1475 \ & \ 0.0021 \\
0.35 \ \ & \ \ 2.5 \ \ & \ \ 0.010  \ \
                         & \ \ 0.2522 \ & \ 0.0025 \ \
                         & \ \ 0.2453 \ & \ 0.0025 \\
0.45 \ \ & \ \ 3.0 \ \ & \ \ 0.010  \ \
                         & \ \ 0.3416 \ & \ 0.0034 \ \
                         & \ \ 0.3414 \ & \ 0.0034 \\
0.55 \ \ & \ \ 3.7 \ \ & \ \ 0.015  \ \
                         & \ \ 0.4308 \ & \ 0.0065 \ \
                         & \ \ 0.4384 \ & \ 0.0066 
\\ \hline
\end{tabular}
\end{center}
\end{table}

\begin{figure}[b]
\begin{center}
        \includegraphics*[width=75mm]{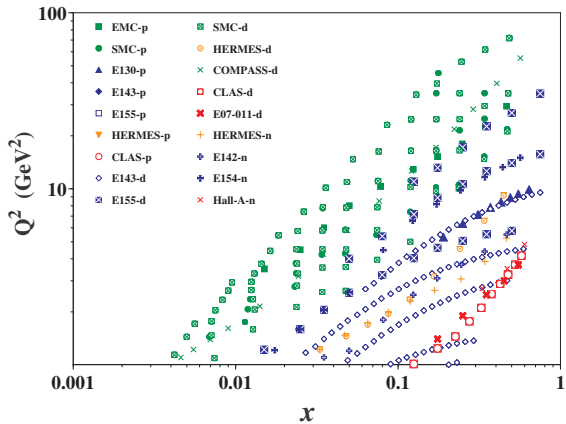} 
\end{center}
\caption{\label{fig:x-Q2}
Kinematical region of the DIS data is shown by $x$ and $Q^2$.
The notations $p$, $n$, and $d$ indicate proton, neutron, and
deuteron, respectively.
}
\end{figure}

The kinematical region of the used DIS data for proton, neutron,
and deuteron is shown in Fig. \ref{fig:x-Q2} by $x$ and $Q^2$.
The CERN data (EMC, SMC, COMPASS) cover a wide $x$ region with
relatively large $Q^2$. The are especially important for determining
the polarized PDFs at small $x$ ($x \sim 0.01$). 
The SLAC (E130, E142, E143, E154, E155), DESY (HERMES), and 
JLab (Hall-A, CLAS, E07-011) data are at relatively large $x$
with small $Q^2$. The are especially important for determining
the polarized PDFs at large $x$ and also possibly the polarized
gluon distribution in comparison with the CERN data.

The details of our analysis method are explained in our previous article
\cite{aac06}, so that only a brief outline is explained in the following.
The polarized PDFs are given at the initial $Q^2$ scale ($\equiv Q_0^2$),
where it is taken $Q_0^2=1$ GeV$^2$, with a number of parameters:
\begin{equation}
   \Delta f(x,Q_0^2) 
   = [\delta x^{\nu}-\kappa (x^{\nu}-x^{\mu})] f(x,Q_0^2) \, ,
\label{eqn:df}
\end{equation}
where $\delta$, $\kappa$, $\nu$, and $\mu$ are parameters to be determined
by a $\chi^2$ analysis, and $f(x)$ is the corresponding unpolarized PDF
\cite{GRV98}. 
Flavor symmetric antiquark distributions are assumed for the polarized PDFs
at the initial scale although the unpolarized antiquark distributions are
not flavor symmetric \cite{flavor3}. 
They are evolved to experimental $Q^2$ points by 
the DGLAP (Dokshitzer-Gribov-Lipatov-Altarelli-Parisi) equations
\cite{dglap}.

The parameters are then determined by a $\chi^2$ analysis of
experimental data on spin asymmetries so as to minimize
\begin{equation}
\chi^2=\sum_i \frac{[ A_i^{\rm data}(x,Q^2)
                     -A_i^{\rm calc}(x,Q^2) ]^2}
                {[\Delta A_i^{\rm data}(x,Q^2) ]^2} .
\label{eqn:chi2}
\end{equation}
Here, the spin asymmetries are $A_1$ in the DIS
and $A_{LL}^{\pi^0}$ in the $\pi^0$ production.
The analyses are done in the next-to-leading order (NLO) of the running
coupling constant $\alpha_s$, and the modified minimal subtraction
($\overline {\rm MS}$) scheme is used. The fragmentation functions
of the HKNS07 (Hirai, Kumano, Nagai, Sudoh in 2007) \cite{hkns07} are used
for describing the fragmentation into the pion. 
However, uncertainties of the fragmentation functions are not included
in this analysis.
The uncertainties of the polarized PDFs are estimated by the Hessian
method:
\begin{equation}
\! \! \! 
        [\delta F(x)]^2=\Delta \chi^2 \sum_{i,j}
          \left( \frac{\partial F(x,a)}{\partial a_i}  \right)_{a=\hat a}
          H_{ij}^{-1}
          \left( \frac{\partial F(x,a)}{\partial a_j}  \right)_{a=\hat a} ,
        \label{eq:erroe-M}
\end{equation}
where $F(x)$ is a polarized PDF, $H_{ij}$ is the Hessian,
the parameters are denoted $a_i$ ($i$=1, 2, ..., $N$),
and $\hat a$ indicates the parameter set
at the minimum $\chi^2$ point.
The value of $\Delta \chi^2$ is taken as $\Delta \chi^2=12.65$
so that the uncertainty indicates the one-$\sigma$-error range 
for the normal distribution in the eleven-parameter space \cite{aac,aac06}.
The details of the analysis conditions and 
the uncertainty-estimation method are found in Ref. \cite{aac06}.

We comment on our choice of $\Delta \chi^2$.
As explained in Ref. \cite{del-chi2}, $\Delta \chi^2 \sim N$ gives
one-$\sigma$ error for a multivariate normal distribution of fit parameters.
We took this $\Delta \chi^2$ value for estimating the uncertainties
by considering it as an appropriate quantity for showing errors of
overall functional behavior. However, this choice is not unique in
global analyses of PDFs \cite{compass-07,del-chi2-pdfs}.
For example, $\Delta \chi^2=1$ is chosen in the COMPASS analysis
for $\Delta g(x)$ \cite{compass-07}.
The difference between the choices,
$\Delta \chi^2 \sim N$ and $\Delta \chi^2 =1$, comes from the fact that
the likelihood function in the multiparameter space is different
from the likelihood function for a physics quantity, for example,
$A_1 (x,Q^2)$ at a fixed kinematical point of $x$ and $Q^2$.
If the latter likelihood is plotted as a function of this physical
quantity (one degree of freedom) $A_1 (x,Q^2)$,
the one-$\sigma$ (68\%) range as the statistical error of the function
corresponds to the $\Delta \chi^2=1$ region. 
However, our experience indicates that the choice $\Delta \chi^2=1$
could be an underestimation of the uncertainties
in comparison with experimental errors and their variations
(see Figs. 1 and 3 in Ref. \cite{aac06}).
Therefore, CTEQ (Coordinated Theoretical/Experimental Project on QCD 
Phenomenology and Tests of the Standard Model) and
MRST  (Martin-Roberts-Stirling-Thorne) use larger $\Delta \chi^2$ values 
( $\Delta \chi^2=50$ or 100) \cite{del-chi2-pdfs} as an effective way
to show the error range. Our choice $\Delta \chi^2 \sim N$ is
a conservative estimation of the uncertainty range by 
considering error correlations among the parameters.
If one would like to use the choice $\Delta \chi^2 = 1$,
one may scale down our polarized PDF uncertainties by 
$1/\sqrt{\Delta\chi^2}$.


\section{Results}
\label{results}



\begin{table*}[t]
\caption{\label{T:chi2}
Numbers of data and $\chi^2$ values are listed for the three analysis sets
with the data in Table \ref{Tabl:JLabD}. The positive and node indicate
a positive distribution and a node-type one, respectively, for $\Delta g(x)$.
The notations $p$, $n$, and $d$ indicate proton, neutron, and deuteron, 
respectively. }
\begin{tabular}{cccccccc} 
\hline
Data set 
               & \ No. of \  
                            & \multicolumn{6}{c}{$\chi^2$} \\
               &  data      & \multicolumn{2}{c}{Set A} 
                            & \multicolumn{2}{c}{Set B} 
                            & \multicolumn{2}{c}{Set C}   \\
               &            & \ \ \ Positive \ \ \ & \ \ \ Node \ \ \ 
                            & \ \ \ Positive \ \ \ & \ \ \ Node \ \ \ 
                            & \ \ \ Positive \ \ \ & \ \ \ Node \ \ \ \\
\hline
$\! \! \! \!$
EMC ($p$)    $\! \! \! \!$  & \  10 & \ 5.17 & $\! \! \! \!$ \ 4.52 
                            & \ 5.18 & $\! \! \! \!$ \ 4.57 
                            & \ 5.17 & $\! \! \! \!$ \ 4.52 $\! \! \! \!$ \\
$\! \! \! \!$                   
SMC ($p$)    $\! \! \! \!$  & \  59 &  56.34 &  $\! \! \! \!$ 53.79 
                            &  56.77 &  $\! \! \! \!$ 53.95
                            &  56.34 &  $\! \! \! \!$ 53.79 $\! \! \! \!$ \\
$\! \! \! \!$
E130 ($p$)  $\! \! \! \!$   & \ \ 8 & \ 5.13 & $\! \! \! \!$ \ 4.84 
                            & \ 5.23 & $\! \! \! \!$ \ 4.97 
                            & \ 5.13 & $\! \! \! \!$ \ 4.84 $\! \! \! \!$ \\
$\! \! \! \!$
E143 ($p$)  $\! \! \! \!$   & \  81 &  59.97 &  $\! \! \! \!$ 59.78 
                            &  60.43 &  $\! \! \! \!$ 60.30 
                            &  59.97 &  $\! \! \! \!$ 59.78 $\! \! \! \!$ \\
$\! \! \! \!$
E155 ($p$)  $\! \! \! \!$   & \  24 &  33.07 &  $\! \! \! \!$ 27.38 
                            &  31.76 &  $\! \! \! \!$ 26.19
                            &  33.06 &  $\! \! \! \!$ 27.39 $\! \! \! \!$ \\
$\! \! \! \!$
HERMES ($p$) $\! \! \! \!$  & \ \ 9 & \ 3.58 & $\! \! \! \!$ \ 3.69 
                            & \ 3.59 & $\! \! \! \!$ \ 3.68 
                            & \ 3.58 & $\! \! \! \!$ \ 3.69 $\! \! \! \!$ \\
$\! \! \! \!$
CLAS ($p$)  $\! \! \! \!$   & \  10 &  14.34 &  $\! \! \! \!$ 15.11 
                            &  17.81 &  $\! \! \! \!$ 21.65 
                            &  14.32 &  $\! \! \! \!$ 15.10 $\! \! \! \!$ \\
\hline                                                        
$\! \! \! \!$
SMC ($d$)  $\! \! \! \!$    & \  65 &  57.39 & $\! \! \! \!$ 56.65 
                            &  57.82 & $\! \! \! \!$ 56.86 
                            &  57.39 & $\! \! \! \!$ 56.65 $\! \! \! \!$ \\
$\! \! \! \!$
E143 ($d$) $\! \! \! \!$    & \  81 &  91.15 & $\! \! \! \!$ 91.86 
                            &  89.84 & $\! \! \! \!$ 89.85 
                            &  91.16 & $\! \! \! \!$ 91.87 $\! \! \! \!$ \\
$\! \! \! \!$
E155 ($d$)  $\! \! \! \!$   & \  24 &  18.60 & $\! \! \! \!$ 22.44 
                            &  18.53 & $\! \! \! \!$ 23.32 
                            &  18.60 & $\! \! \! \!$ 22.44 $\! \! \! \!$ \\
$\! \! \! \!$
HERMES ($d$) $\! \! \! \!$  & \ \ 9 &  11.19 & $\! \! \! \!$ \ 8.10  
                            &  11.93 & $\! \! \! \!$ \ 7.65 
                            &  11.20 & $\! \! \! \!$ \ 8.10 $\! \! \! \!$\\
$\! \! \! \!$
COMPASS ($d$) $\! \! \! \!$ & \  15 & \ 8.73 & $\! \! \! \!$ 10.49 
                            & \ 9.30 & $\! \! \! \!$ 11.89 
                            & \ 8.73 & $\! \! \! \!$ 10.49 $\! \! \! \!$\\
$\! \! \! \!$
CLAS ($d$)   $\! \! \! \!$  & \  10 & \ 5.73 & $\! \! \! \!$ \ 4.76 
                            & \ 5.55 & $\! \! \! \!$ \ 5.53 
                            & \ 5.73 & $\! \! \! \!$ \ 4.76 $\! \! \! \!$ \\
$\! \! \! \!$
E07-011 ($d$) $\! \! \! \!$ & \ \ 5 &  $-$   &  $\! \! \! \!$ $-$ 
                            &  $-$   & $\! \! \! \!$ $-$  
                            & \ 0.00 & $\! \! \! \!$ \ 0.00 $\! \! \! \!$ \\
\hline                                                        
$\! \! \! \!$
E142 ($n$)  $\! \! \! \!$   & \ \ 8 & \ 2.60 & $\! \! \! \!$ \ 2.48 
                            & \ 2.78 & $\! \! \! \!$ \ 2.68 
                            & \ 2.60 & $\! \! \! \!$ \ 2.48 $\! \! \! \!$ \\
$\! \! \! \!$
E154 ($n$)  $\! \! \! \!$   & \  11 & \ 3.32 & $\! \! \! \!$ \ 4.09 
                            & \ 3.07 & $\! \! \! \!$ \ 3.12 
                            & \ 3.32 & $\! \! \! \!$ \ 4.10 $\! \! \! \!$ \\
$\! \! \! \!$
HERMES ($n$) $\! \! \! \!$  & \ \ 9 & \ 2.26 & $\! \! \! \!$ \ 2.41
                            & \ 2.18 & $\! \! \! \!$ \ 2.36 
                            & \ 2.26 & $\! \! \! \!$ \ 2.41 $\! \! \! \!$ \\
$\! \! \! \!$
Hall-A ($n$) $\! \! \! \!$    & \ \ 3 & \ 3.09 & $\! \! \! \!$ \ 3.49
                            & \ 2.89 & $\! \! \! \!$ \ 2.70
                            & \ 3.09 & $\! \! \! \!$ \ 3.48 $\! \! \! \!$ \\
\hline                                                          
$\! \! \! \!$
DIS total  $\! \! \! \!$    &   441 & 381.66  & $\! \! \! \!$ 375.87  
                            & 384.65  & $\! \! \! \!$ 381.27  
                            & 381.66  & $\! \! \! \!$ 375.87 $\! \! \! \!$ \\
$\! \! \! \!$
PHENIX ($\pi^0$) $\! \! \! \!$ & \  10 & \   $-$  & $\! \! \! \!$ \   $-$  
                            & \ 12.43  & $\! \! \! \!$ \  11.32  
                            & \   $-$  & $\! \! \! \!$ \   $-$  $\! \! \! \!$ \\
\hline                                                          
$\! \! \! \!$
Total  $\! \! \! \!$        &   451 & 381.66  & $\! \! \! \!$ 375.87  
                            & 397.08  & $\! \! \! \!$ 392.60  
                            & 381.66  & $\! \! \! \!$ 375.87 $\! \! \! \!$ \\
$\! \! \! \!$
($\chi^2$/d.o.f.) $\! \! \! \!$  &       & (0.90)   & $\! \! \! \!$ (0.88) 
                            & (0.91)   & $\! \! \! \!$ (0.90)  
                            & (0.89)   & $\! \! \! \!$ (0.87) $\! \! \! \!$  \\
\hline
\end{tabular}
\end{table*}


\begin{table*}[t]
\caption{\label{T:1st-moments}
First moment of the polarized gluon distribution function
and quark spin content
$\Delta \Sigma=\sum_i \int_0^1 dx \, [\Delta q_i(x)+\Delta\bar q_i(x)]$
at $Q^2=1$ GeV$^2$.}
\vspace{2mm}
\begin{tabular}{ccccccc} 
\hline
            &  \multicolumn{2}{c}{Set A} 
            &  \multicolumn{2}{c}{Set B} 
            &  \multicolumn{2}{c}{Set C}   \\
            &  \ \ \ Positive \ \ \ & \ \ \ Node \ \ \ 
            &  \ \ \ Positive \ \ \ & \ \ \ Node \ \ \ 
            &  \ \ \ Positive \ \ \ & \ \ \ Node \ \ \ \\
\hline   
$\! \! \! \!$                                                         
$\Delta \Sigma$ 
           \ \ & \ \ \ \ 0.24 \ $\pm$ 0.07 \ \ & \ \ 0.22 $\pm$ 0.08 \ \ \ 
               & \ \ \ \ 0.26 \ $\pm$ 0.06 \ \ & \ \ 0.25 $\pm$ 0.07 \ \ \ 
               & \ \ \ \ 0.24 \ $\pm$ 0.05 \ \ & \ \ 0.22 $\pm$ 0.05 \ \ \ \\
$\! \! \! \!$
$\Delta G$ \ \ & \ \ \ \ 0.63 \ $\pm$ 0.81 \ \ & \ \ 0.94 $\pm$ 1.66 \ \ \ 
               & \ \ \ \ 0.40 \ $\pm$ 0.28 \ \ & \ $-$0.12 $\pm$ 1.78 \ \ \ \  
               & \ \ \ \ 0.63 \ $\pm$ 0.45 \ \ & \ \  0.94 $\pm$ 1.09 \ \ \ \\
\hline
\end{tabular}
\end{table*}

\begin{table}[b]
\caption{The first moment $\Delta G$ and its uncertainty
         $\delta \Delta G$ at $Q^2=1$ GeV$^2$ in the range $0.1<x<1$.
         Here, the integrals are calculated in the limited $x$ range by
         $\Delta G (x>0.1)\equiv\int_{0.1}^1 dx \Delta g(x)$.}
\label{T:1st-moment-01}
\vspace{2mm}
\begin{center}
\begin{tabular}{ccccc}
\hline
\ Function \ & \ Set \
             &   $\Delta G (x>0.1)$         \ 
             &   $\delta \Delta G (x>0.1)$  \
             &   $\frac{\delta \Delta G (x>0.1)}{\Delta G (x>0.1)}$ \ \\
\hline
Positive     & A   & 0.53   & 0.72   & 1.36 \\
             & B   & 0.36   & 0.26   & 0.71 \\
             & C   & 0.53   & 0.38   & 0.73 \\
\hline
Node         & A   & 0.87   & 0.89   & 1.02 \\
             & B   & 0.40   & 0.31   & 0.77 \\
             & C   & 0.87   & 0.47   & 0.54 \\
\hline
\end{tabular}
\end{center}
\end{table}

Three types of analyses are done with the different data sets
in Table \ref{Tabl:JLabD}. The $\chi^2$ values are shown for each
analysis in Table \ref{T:chi2}. The total $\chi^2$ values per degrees of
freedom are in the range $\chi^2$/d.o.f.=0.87$-$0.91, so that all
the analyses are successful in explaining the data. 
The $\chi^2$ value for the expected E07-011 data
in the analysis C is very small ($\sim 10^{-4}$) because the values
of the data are taken from the analysis-A results and there are
no statistical variations. The polarized PDFs of the analysis C
should be close to the ones of the analysis A.
Comparison between A and C results is useful for discussing
constraints from the E07-011 data on $\Delta g(x)$ determination
by noting differences between their uncertainties.
First moments of the polarized PDFs are listed in Table \ref{T:1st-moments}.
The first moments of $\Delta u_v(x)$ and $\Delta d_v(x)$ are fixed by
semileptonic decays: $\int_0^1 dx \, \Delta u_v(x)=0.926$
and $\int_0^1 dx \, \Delta d_v(x)=-0.341$  \cite{aac}.
In order to discuss constraints from the RHIC-$\pi^0$ and JLab-E07-011 
data on the gluon distribution at relatively large $x$, we also show
the ``first moment" calculated by the integral
$\Delta G(x>0.1)=\int_{0.1}^1 dx g(x)$ in Table \ref{T:1st-moment-01}.

\subsection{Impact of RHIC $\pi^0$ data}
\label{rhic-pi0}

\begin{figure}[b]
\begin{center}
     \includegraphics*[width=40mm]{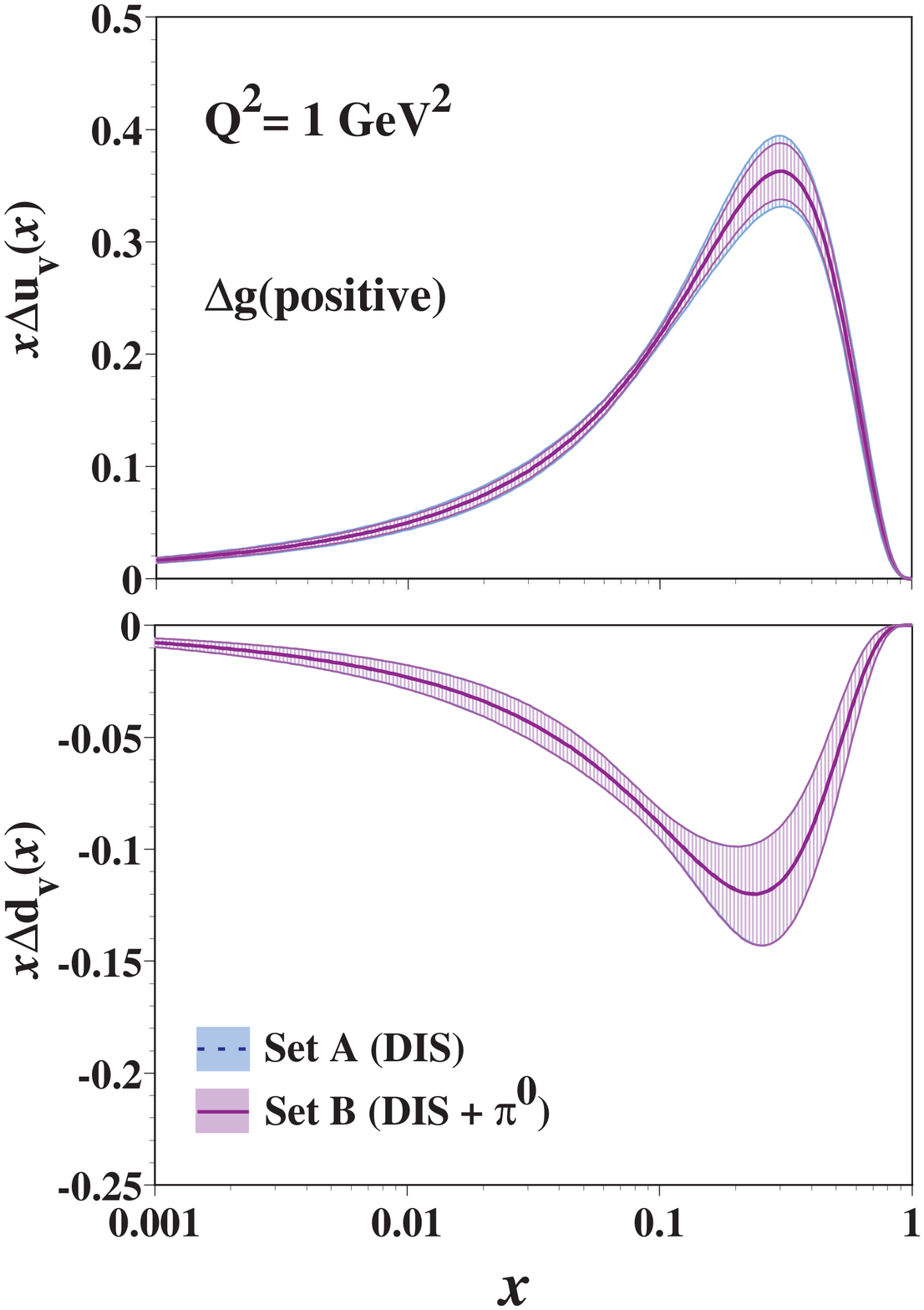} 
     \hspace{1mm}
     \includegraphics*[width=40mm]{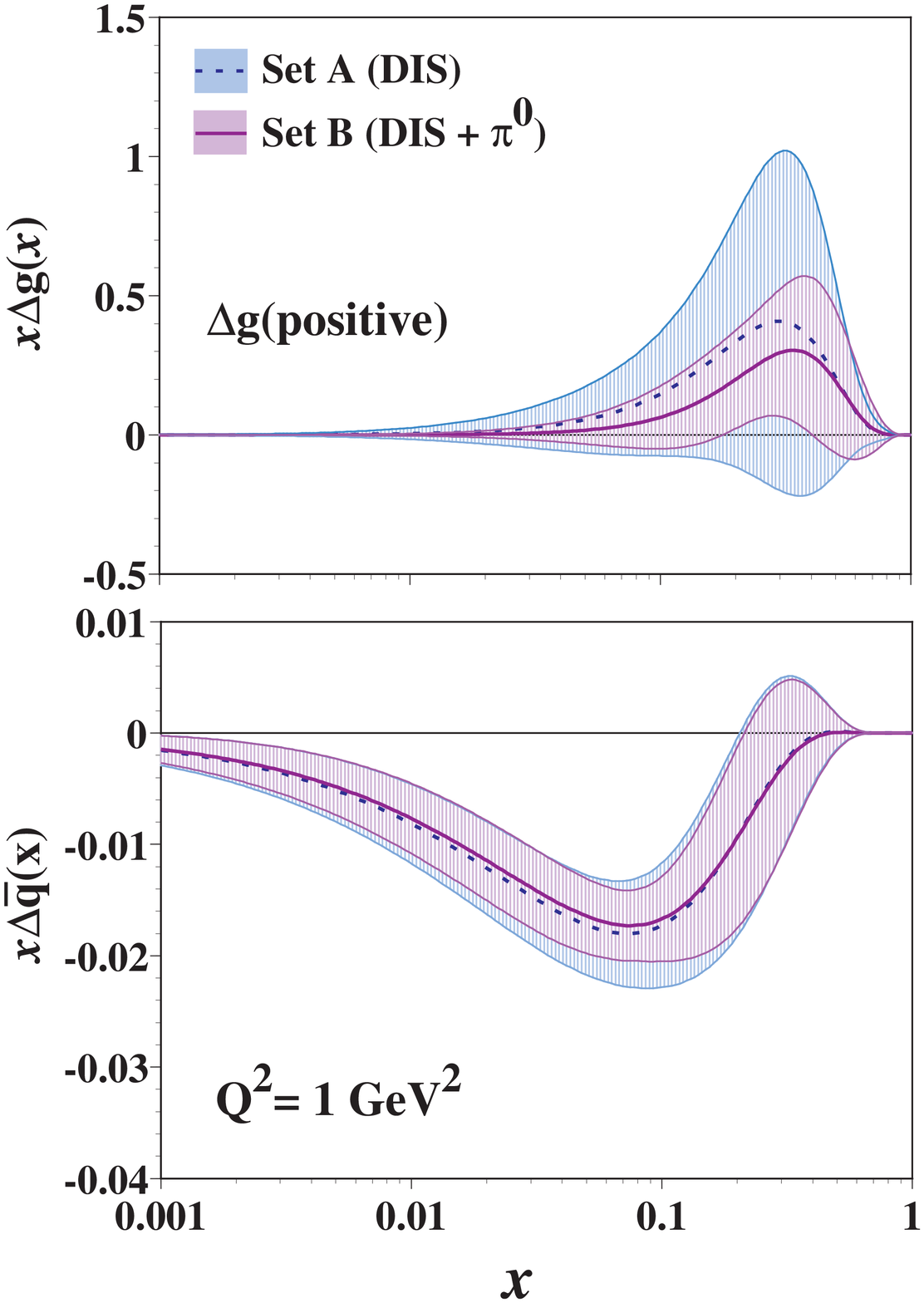}
\end{center}
\caption{\label{fig:setAB}
Comparison of the polarized PDFs between the analyses A and B. 
They are shown by the dashed and solid curves for
the analyses A and B, respectively, at $Q^2$=1 GeV$^2$.
Their uncertainties are shown by the shaded bands.
}
\end{figure}

First, effects of the RHIC $\pi^0$ data \cite{phenix-pi}
are shown in comparison with the polarized PDFs obtained only 
by the DIS data.
Using the preliminary PHENIX $\pi^0$ data of the RHIC RUN-5,
we have already shown their impact on the gluon distribution
in Ref. \cite{aac06}. The current analysis uses the published RUN-5 data.
Analysis-B results are compared with analysis-A ones
in Fig. \ref{fig:setAB}, where the polarized PDFs
$x\Delta u_v$, $x\Delta d_v$, $x\Delta \bar q$, and $x\Delta g$ 
and their uncertainties are shown at $Q^2$=1 GeV$^2$.
The set A uses only the DIS data and the set B also includes
the $\pi^0$ data as listed in Table \ref{Tabl:JLab-E07-011}.

The polarized valence-quark distributions $\Delta u_v(x)$ and
$\Delta d_v(x)$ are not changed even if the RHIC $\pi^0$ data 
are included in the analysis. However, there are significant
effects on the other distributions, especially on the polarized
gluon distribution. The gluon distribution has a huge uncertainty
if it is determined only by the DIS data (analysis A); however,
the uncertainty band becomes significantly small in the analysis B
because of the addition of the $\pi^0$ data.
It is known that gluon-gluon interaction subprocesses dominate
the $\pi^0$-production cross section especially at small $p_T$,
so that the $\pi^0$ data are sensitive to the gluon distribution.
We notice that the antiquark uncertainty is slightly reduced
in the set B and it is caused by the error correlation between
the antiquark and gluon distributions as pointed out
in Ref. \cite{aac06}.

The double spin asymmetry for the $\pi^0$ production indicates
a negative value at $p_T=2.38$ GeV. It suggests a node
type distribution which vanishes at $x \sim 0.1$ as investigated
in Refs. \cite{aac06,dssv08,negativeAsym}. We also made a global 
analysis by assigning an initial distribution to a node-type one
for $\Delta g(x)$. 
Minimum $\chi^2$ values per degrees are slightly smaller
in the node-type results than the values in the positive ones.
In the node-type results of the analysis A, the gluon distribution
is positive at $x>0.2$ and is slightly negative at small $x$.
The negative gluon distribution at $x<0.1$ becomes smaller
in the node-type of the analysis B.

As noticed in Ref. \cite{aac06}, the differences between
HERMES ($Q^2 \simeq 1$ GeV$^2$) and COMPASS ($Q^2 \simeq 6$ GeV$^2$)
data of $g_1^d$ at $x \sim 0.05$ could suggest a positive
$\Delta g(x)$ at large $x$ ($>0.2$) although it may be explained
by a higher-twist effect \cite{lss06}. The gluon distribution contributes
to $g_1(x)$ as a NLO effect and the differences could
be explained by this NLO effect with $\Delta g(x)>0$ at $x>0.2$.
In fact, all our global analyses indicate positive distributions 
at large $x$.

\begin{figure}[b]
\begin{center}
        \includegraphics*[width=50mm]{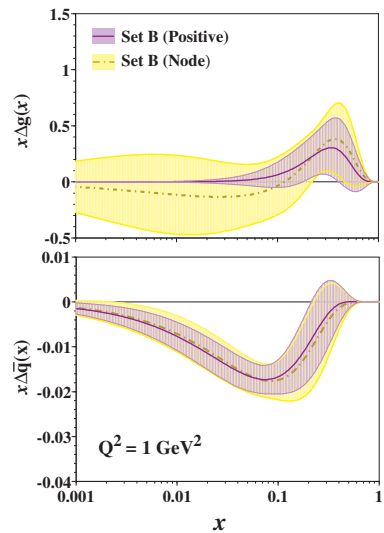} 
\end{center}
\caption{\label{fig:gqbar-posi-node-setB}
Polarized antiquark and gluon distributions by the analysis B
which includes the RHIC $\pi^0$-production data.
The distributions and their uncertainties are calculated
at $Q^2=1$ GeV$^2$.
Positive and node-type solutions are shown by the solid
and dashed-dot curves, respectively.
}
\end{figure}

The node-type distributions are compared with the positive ones
at $Q^2 =1$ GeV$^2$ in Fig. \ref{fig:gqbar-posi-node-setB}.
Both distributions are determined by the set-B analysis.
It is clear from this figure that the gluon
distribution at $x<0.1$ cannot be determined by the current data
including the RHIC $\pi^0$ data. In fact, the uncertainty of $\Delta G$
is large in the node-type distribution and it is obvious that
a large contribution comes from the small $x$ region.
Because there is no constraint at small $x$, uncertainty
reduction of $\Delta G$ is not very obvious due to the $\pi^0$ data
from the set A to set B
as shown in Table \ref{T:1st-moments}.

In order to investigate a possible improvement on $\Delta G$,
we discuss a contribution to the first moment from
the region of $x>0.1$ in Table \ref{T:1st-moment-01}.
The relative uncertainties become smaller in the set B from the set A:
$\delta\Delta G(x>0.1)/\Delta G(x>0.1)$
= 1.36 $\rightarrow$ 0.71 in the positive distribution and 
  1.02 $\rightarrow$ 0.77 in the node one.
There are significant reductions in the uncertainties (24\%$\sim$47\%),
which suggests that the $\pi^0$ data impose a significant constraint
in the gluon distribution at $x>0.1$ in addition to the aforementioned
DIS constraint from the scaling violation. 

\begin{figure}[b]
\begin{center}
     \includegraphics*[width=40mm]{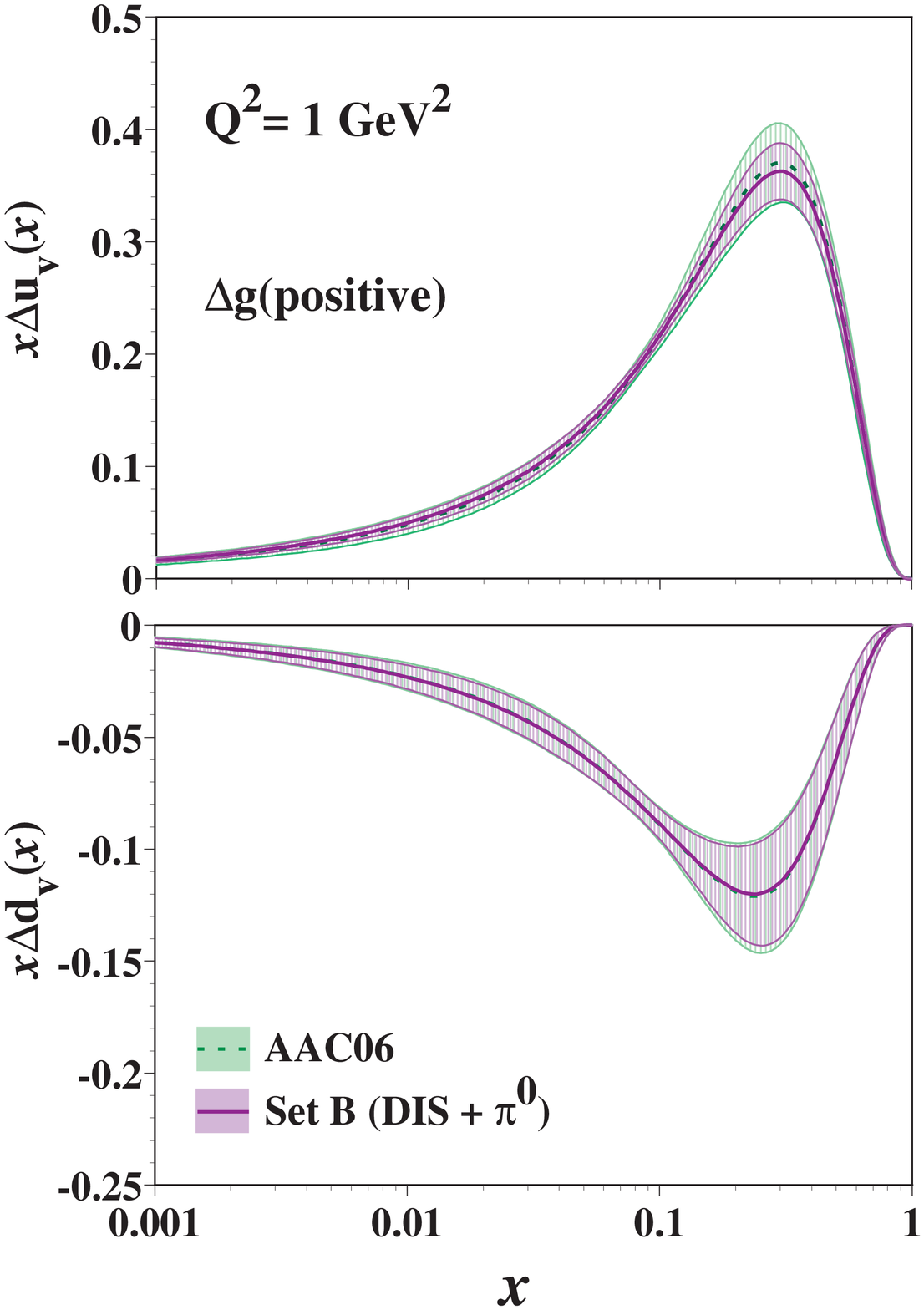} 
     \hspace{1mm}
     \includegraphics*[width=40mm]{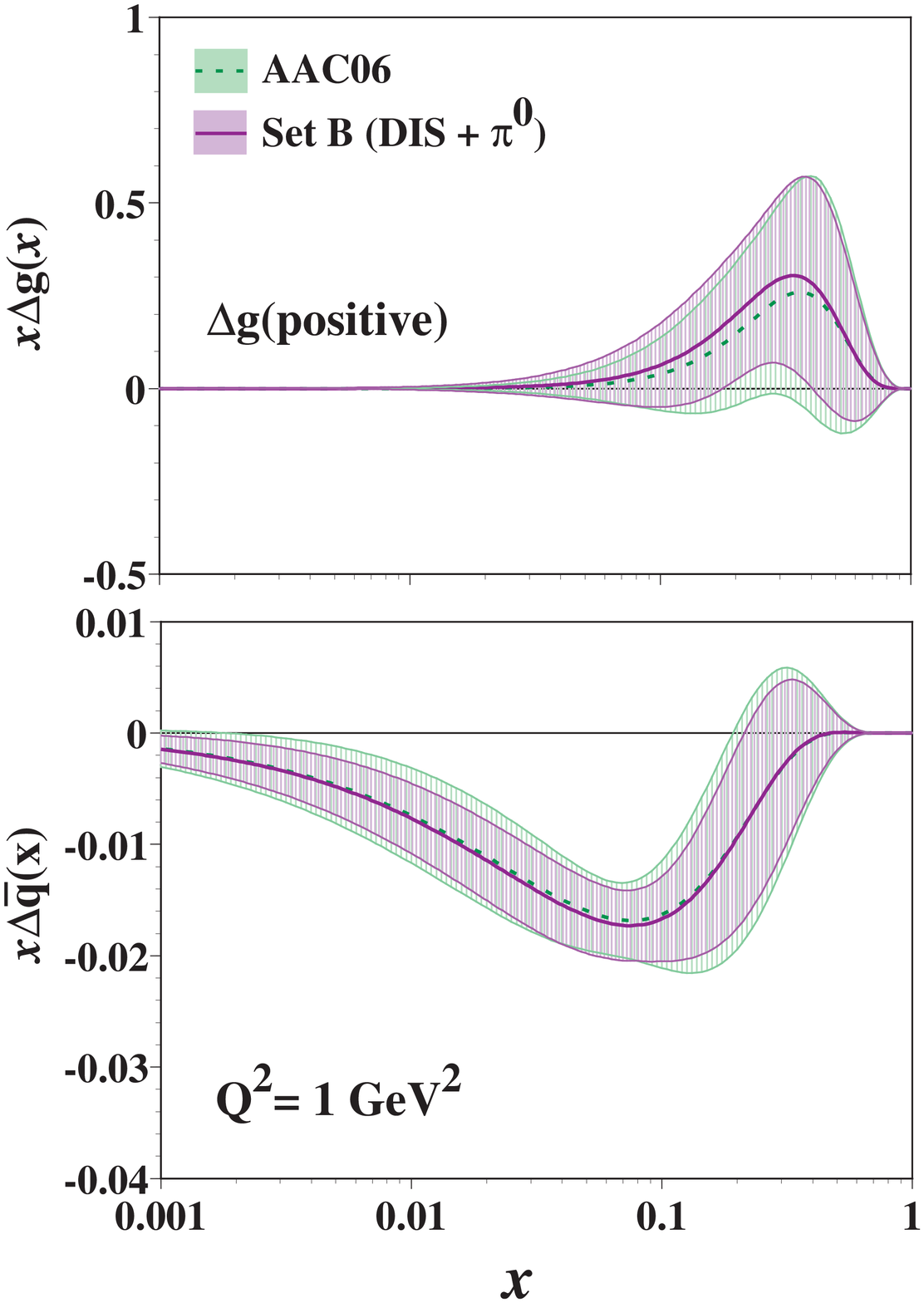}
\end{center}
\caption{\label{fig:comp-aac06-aac08}
Comparison of the polarized PDFs between the current analysis B
and AAC06 for the positive $\Delta g(x)$. 
They are shown by the dashed and solid curves for the AAC06
and the current analysis (AAC08), respectively, at $Q^2$=1 GeV$^2$.
Their uncertainties are shown by the shaded bands.
}
\end{figure}

We discuss a comparison with our previous work of AAC06 \cite{aac06}.
The positive distributions of the set B and their uncertainties
are compared with the AAC06 results in Fig. \ref{fig:comp-aac06-aac08}.
There are differences between the used data sets. As explained in
Sec. \ref{intro}, additional data, which are not used in the AAC06
analysis, are those of the CLAS \cite{clas-06}
and COMPASS \cite{compass-07} collaborations. The PHENIX $\pi^0$
data \cite{phenix-pi} were also slightly changed from the preliminary
ones at the stage of the AAC06 publication. The used fragmentation
functions are also changed from the functions by KKP
(Kniehl, Kramer, and P\"otter) for the ones by the HKNS07. 

As shown in Fig. \ref{fig:comp-aac06-aac08}, the distributions and
their uncertainties are almost the same. There are slight improvements
in the valence-quark and antiquark distributions, which
is manly due to the accurate measurements at medium $x$ by the
CLAS and COMPASS collaborations. The slight reduction in
the $\Delta g(x)$ uncertainty band should come from the error
correlation between the quark and gluon distributions. 
In addition, there are effects due to changes in
the $\pi^0$ data from the preliminary version and 
in the fragmentation functions. However, these effects are not
significant because the distributions themselves are almost 
the same. We also compared the AAC06 and current distributions
for the node type, but they are also almost the same
in both distributions and uncertainties.

\begin{figure}[t]
\begin{center}
        \includegraphics*[width=85mm]{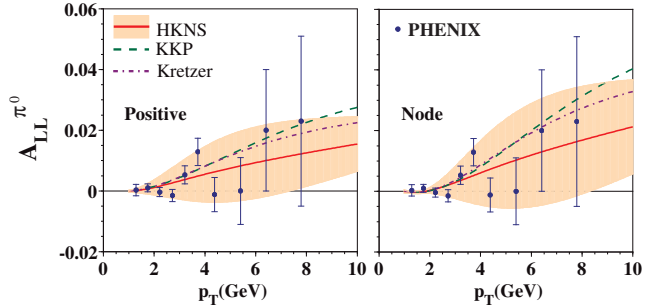} 
\end{center}
\caption{\label{fig:pi-frag}
Set-B results are compared with the PHENIX data
in the longitudinal double spin asymmetry
of $\vec p +\vec p \rightarrow \pi^0 +X$ \cite{phenix-pi}.
The solid curves and the shaded bands indicate
theoretical asymmetries and their uncertainties by the HKNS07
fragmentation functions. The results by the KKP and Kretzer 
fragmentation functions are shown by the dashed and dashed-dot
curves, respectively.
}
\end{figure}

It should be noted that the description of a $\pi^0$ production
cross section depends much on pion fragmentation functions. 
Here, the HKNS07 functions \cite{hkns07} are employed in our global
analyses for describing the pion cross section. As noted in
Ref. \cite{hkns07}, there are large uncertainties in the fragmentation
functions, especially in the gluon and light-quark functions.
Effects of such uncertainties are shown in Fig. \ref{fig:pi-frag}
together with the RHIC $\pi^0$-production data and
the asymmetries calculated by using the fragmentation functions of
Kretzer and KKP as other examples. 

The pion production is sensitive to the gluon fragmentation function,
which has a large uncertainty because it has been determined mainly
by the scaling violation of $e^+ e^- \rightarrow h+X$ data. 
However, we expect that the situation will improve if the Belle and Babar
collaborations provide low-energy $e^+ e^-$ data for clarifying
the scaling violation.
In our analysis of the pion production, the NLO effects are approximately
taken into account as a $K$-factor by first calculating the leading-order
(LO) cross sections. Therefore, the LO fragmentation functions are 
used in the analysis, and the asymmetry has the large uncertainty bands
as shown in the figure. The gluon fragmentation function is determined
more precisely in the NLO \cite{hkns07}, so that the uncertainties 
of $A_{LL}^{\pi^0}$ are reduced about 1/3. In any case, there are
significant effects on the spin asymmetry $A_{LL}^{\pi^0}$ from 
the fragmentation functions, which indicates the importance
to access the uncertainties from the fragmentation functions
in determining the error of $\Delta g(x)$. 

\subsection{Impact of JLab-E07-011 data}
\label{pr-07-011}

\begin{figure}[b]
\begin{center}
     \includegraphics*[width=40mm]{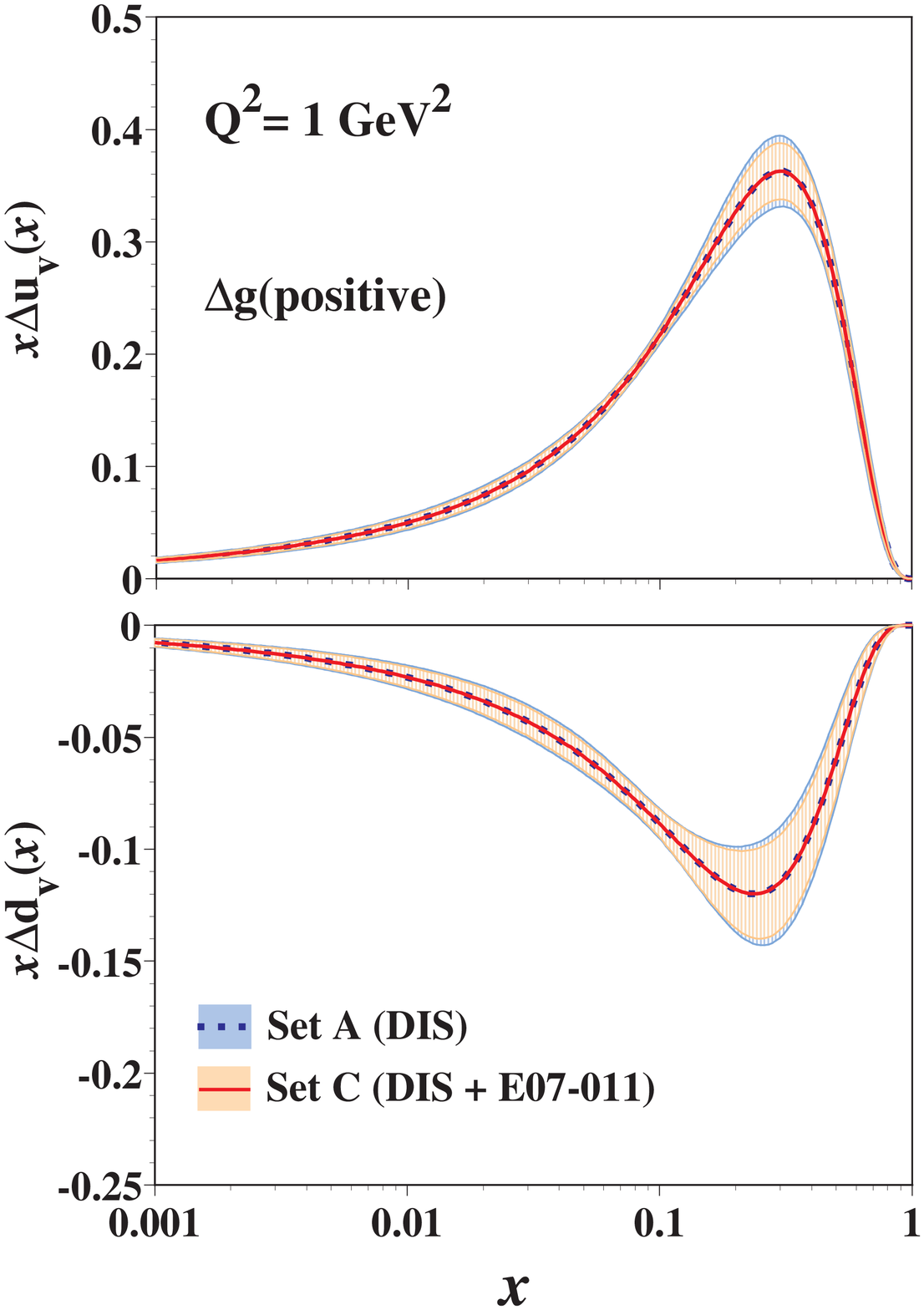} 
     \hspace{1mm}
     \includegraphics*[width=40mm]{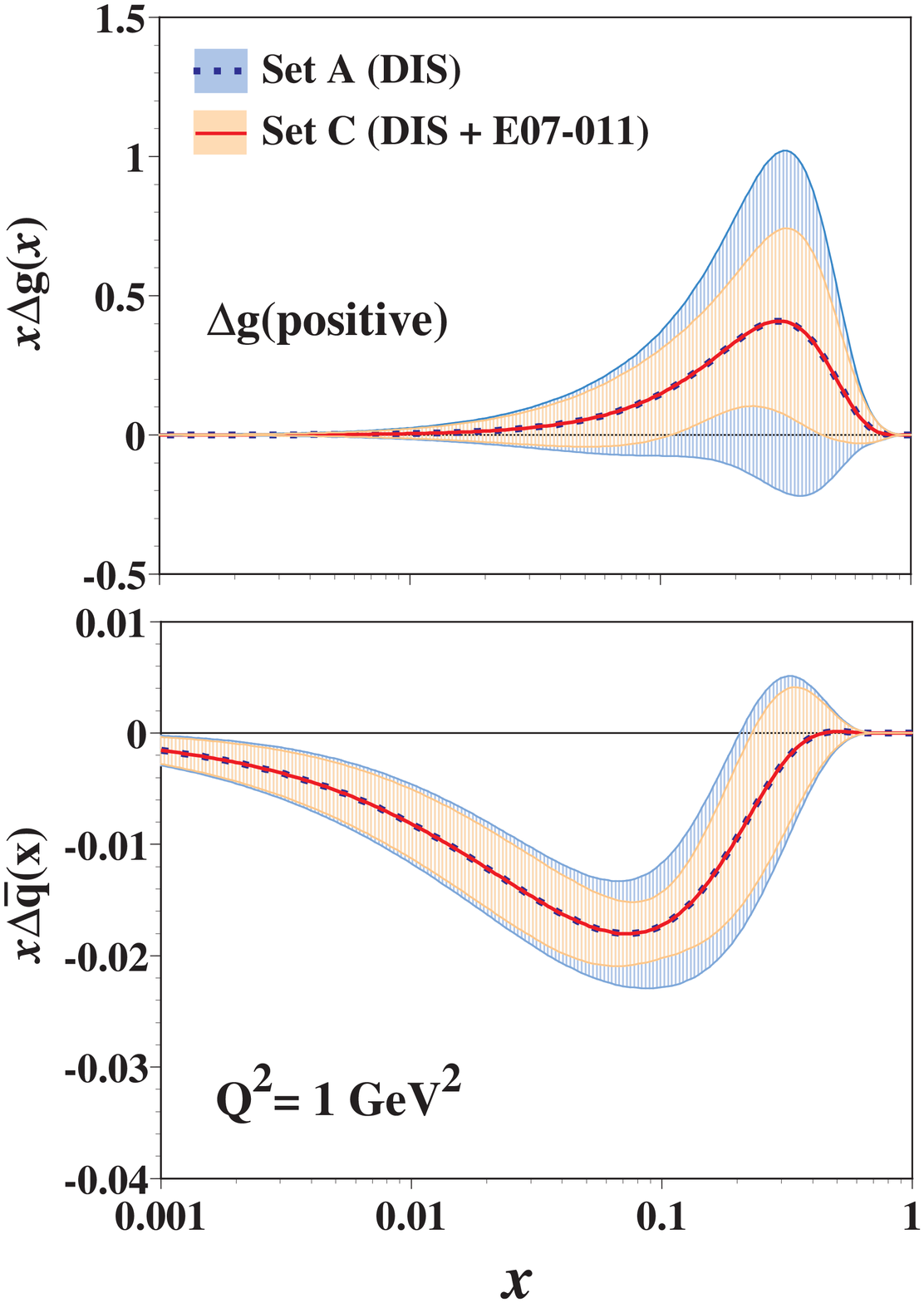}
\end{center}
\caption{\label{fig:setAC}
Comparison of the polarized PDFs between the analyses A and C. 
They are shown by the dashed and solid curves for
the analyses A and C, respectively, at $Q^2$=1 GeV$^2$.
Their uncertainties are shown by the shaded bands.
}
\end{figure}

We discuss the impact of the proposed experiment JLab-E07-011
\cite{jlab-pr-07-011} to measure the structure function $g_1$
for the deuteron on the determination of the polarized PDFs.
The expected E07-011 data \cite{JLabD-ProData}
in Table \ref{Tabl:JLab-E07-011} are included in the analysis C. 
In order to find an improvement to the analysis A with the current DIS data,
the polarized PDFs of the analyses C are compared with the analysis-A
results in Fig. \ref{fig:setAC} at $Q^2$=1 GeV$^2$.
The uncertainties are shown by the shaded bands. 

Both distributions are almost the same because the E07-011
data are assumed to be the same asymmetries obtained
in the analysis A.
The uncertainties are similar in the polarized
valence-quark distributions in Fig. \ref{fig:setAC},
whereas the uncertainty of the antiquark distribution becomes smaller
in the analysis C than the one of the analysis A. 
It is also clear that the uncertainty of the gluon distribution
is significantly reduced in the analysis C.

\begin{figure}[b]
\begin{center}
        \includegraphics*[width=85mm]{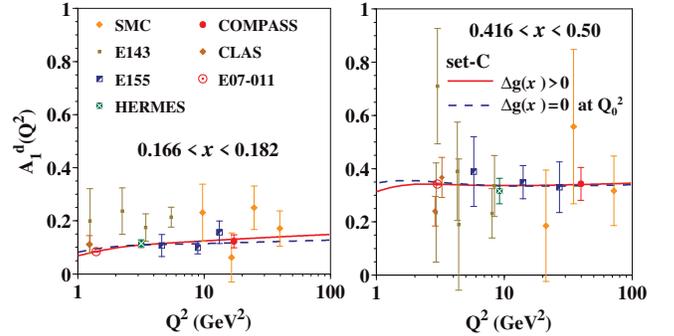} 
\end{center}
\caption{\label{fig:q2-dependence}
$Q^2$ dependence is shown in the spin asymmetry $A_1$ for
the deuteron. The data at $0.166<x<0.182$ and $0.416<x<0.50$
are shown in comparison with the theoretical curves of
the analysis C. The solid curves are calculated by
the set-C distributions with $\Delta g(x)>0$, and the dashed
ones are by the set-C distributions with a different gluon
distribution, $\Delta g(x)=0$ at $Q_0^2=1$ GeV$^2$.
The theoretical curves are calculated at $x=0.175$ and $x=0.45$. 
}
\end{figure}

There is a possibility that the reduction of the gluon uncertainty 
due to the E07-011 data could come from the $Q^2$ dependence
in comparison with other data.
In order to discuss such a possibility, we show the $Q^2$ dependence of
$A_1^d$ within the $x$ range of the E07-011 data in Fig. \ref{fig:q2-dependence}.
Two $x$ ranges are shown. One is at $0.166<x<0.182$ and the other is
at $0.416<x<0.50$. The solid and dashed curves are theoretical set-C results
for the positive $\Delta g(x)$ distribution and an artificial one with
$\Delta g(x)=0$ at the initial scale $Q_0^2$. We notice that differences
between these curves are very small in comparison with the experimental
errors. The results suggest that the current data should not be accurate
enough to probe the polarized gluon distribution from the $Q^2$ dependence
in the $x$ region, $0.1<x<0.5$. Therefore, the uncertainty reduction 
in $\Delta g$ due to the E07-011 data should come from other sources.

There are two possibilities for the reduction of the gluon uncertainty.
First, it could be due to the error correlation between the antiquark
and gluon distributions. A more accurate determination of
antiquark distributions results in the improvement on 
the gluon determination through the error correlation
between polarized antiquark and gluon distributions.

\begin{figure}[b]
\begin{center}
        \includegraphics*[width=70mm]{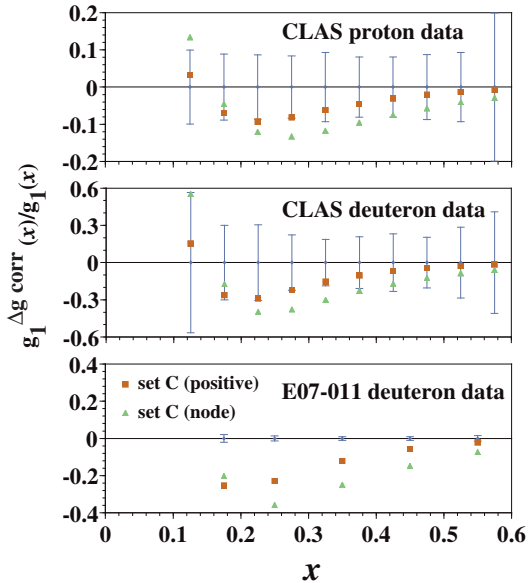} 
\end{center}
\caption{\label{fig:dg_corr_vs_experr}
The ratio of the gluon NLO-correction term to the polarized
structure function $g_1(x,Q^2)$ in Eq. (\ref{eqn:glue-term}).
It is compared with experimental errors, which 
are shown by the ratio
$\delta g_1(x,Q^2)/g_1(x,Q^2)$ in Eq. (\ref{eqn:g1-error}).}
\end{figure}

Second, it could be due to the NLO term with the gluon distribution.
In Fig. \ref{fig:dg_corr_vs_experr}, we show the ratio of the gluon
NLO term to $g_1$:
\begin{equation}
\frac{1}{g_1(x,Q^2)}
\frac{1}{2}\sum\limits_{i=1}^{n_f} e_{i}^2 
\int^{1}_{x} \frac{dz}{z}
\Delta C_g(x/z,Q^2) \Delta g(z,Q^2),
\label{eqn:glue-term}
\end{equation}
where $e_i$ is the quark charge of the flavor $i$,
and $\Delta C_g$ is a gluonic coefficient function.
The ratios are calculated by using the polarized PDFs of the analysis C,
and they are shown by solid boxes and triangles.
The NLO corrections are calculated
at the corresponding experimental $x$ and $Q^2$ values.
The experimental errors are shown by the ratio
\begin{equation}
  \frac{\delta g_1(x,Q^2)}{g_1(x,Q^2)}
      =\frac{\delta A_1^{exp}(x,Q^2)}{A_1^{exp}(x,Q^2)},
\label{eqn:g1-error}
\end{equation}
where $A_1^{exp}(x,Q^2)$ and $\delta A_1^{exp}(x,Q^2)$ are
an experimental spin asymmetry and its error.

It is obvious from Fig. \ref{fig:dg_corr_vs_experr} that the gluonic NLO
effects are within the experimental errors of the current CLAS data
\cite{clas-06} for both proton and deuteron.
However, the proposed experiment E07-011 is very accurate,
and expected errors are much smaller than the NLO effects
shown by the boxes and triangles.
It suggests that the E07-011 measurements should be valuable for
determining not only the quark and antiquark distributions but also
the gluon distribution. However, higher-twist effects become apparent
at small $Q^2$, so that careful consideration is needed also for them
as well as the gluonic NLO contributions if such high-precision data
are taken at JLab.

Another global analysis is made by removing only the CLAS data
from the data set A, and results indicate that the $\Delta g(x)$
uncertainty is slightly reduced. 
From Fig. \ref{fig:dg_corr_vs_experr}, we find that such a reduction
due to the CLAS data is caused mainly by the error correlation
between antiquark and gluon distributions
because the NLO effects are within the experimental errors.
On the other hand, the reduction due to the E07-011 data
in Fig. \ref{fig:setAC} is caused by both the error correlation
and the NLO contributions because the NLO effects are much larger
than the errors.
It is important to find that the E07-011 data are very accurate to
probe the higher-order gluonic effects in the structure function $g_1$.

We comment whether or not such a gluonic NLO term might be effectively
absorbed into variations of polarized antiquark distributions because
it is generally considered to be difficult to separate both distributions
by methods other than the scaling violation. However, it is
not the case in our analysis with the E07-011 data.
As clearly shown in Fig. \ref{fig:setAC}, the reduction of the gluon
uncertainty is much larger than the antiquark one. It cannot be
simply explained by the error correlation between the antiquark
and gluon distributions. It suggests that the E07-011 data should
have a significant impact on the gluon distribution and that 
the effect should not be absorbed into the antiquark distributions. 

\begin{figure}[b]
\begin{center}
        \includegraphics*[width=50mm]{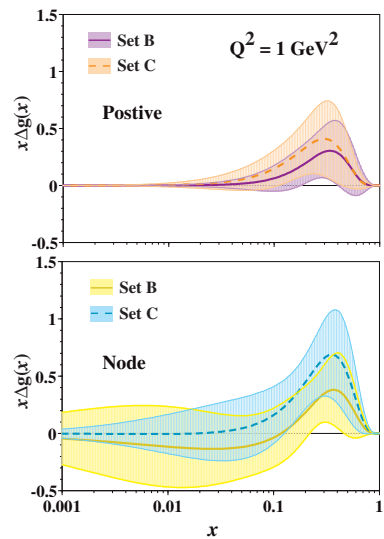} 
\end{center}
\caption{\label{fig:g-setBC}
Polarized gluon distributions and their uncertainties
of the analysis C are compared with the ones of the analysis B
at $Q^2$=1 GeV$^2$. Both positive and node-type distributions
are shown. The set B and C results are shown by the solid and 
dashed curves.
}
\end{figure}

Next, we compare the gluon distributions $\Delta g(x)$ of 
the analysis C with the analysis-B distributions
in Fig. \ref{fig:g-setBC}.
This figure is intended to show the impact of the E07-011 data
in comparison with effects of the RHIC $\pi^0$ data.
We found in Fig. \ref{fig:g-setBC} that both B and C 
distributions are similar. However, it should be noted that
the positivity condition $|\Delta g(x)| \le g(x)$ is not satisfied
for the node-type solution in the analyses B and C at large $x$.
It was difficult to get a converging result within the positivity
condition especially for the parameter $\delta_g$, so that 
the condition is not imposed in the analysis.
We also notice that the relative uncertainties for C are roughly
the same with the ones for B. This fact indicates that the effects
of the E07-011 data on the determination of $\Delta g(x)$ are
comparable with those of the $\pi^0$ data.

Mechanisms for the reductions of the gluon uncertainties
are completely different in both cases. In the analysis B,
the reduction is due to the gluon-gluon scattering subprocesses
in the $\pi^0$ production, and it is due to the error correlation
and the NLO gluon term of $g_1$ in the analysis C.
Nonetheless, it is noteworthy that high-precision lepton-scattering
data can be used for determining the polarized gluon distribution
and that the improvement is as good as the current RHIC $\pi^0$ data.

We comment on possible higher-twist effects.
Because the JLab data are taken at relatively small $Q^2$,
higher-twist corrections may affect the results. 
Such effects are not included in our leading-twist analysis.
There were studies on the higher-twist effects in the structure
function $g_1$ \cite{bb02,lss06}.
We also have done a preliminary analysis on such effects; however,
they are not precisely determined. Namely, uncertainties of
the higher-twist corrections are large, which indicates
that they may not be determined by the current experimental 
data as pointed out in Ref. \cite{bb02}.
If additional parameters are introduced in the higher-twist term, 
they have very large errors, which makes it difficult to get a reliable
polarized PDF set by the current global analysis. We will work on an analysis
for obtaining a possible higher-twist correction from the experimental
data.

Finally, we discuss the first moment of the polarized gluon distribution,
$\Delta G$ $(= \int_0^1 \Delta g(x) dx)$. The first moments and
their uncertainties are shown in Table \ref{T:1st-moments}
for all the analyses. 
We find that the uncertainties become smaller
in the set C from the set A:
$\delta\Delta G$=
0.81 $\rightarrow$ 0.45 in the positive distribution and 
1.66 $\rightarrow$ 1.09 in the node one.
In order to investigate a possible constraint on $\Delta G$
in the large-$x$ region, we also showed the first moment in
the kinematical region of $x>0.1$ in Table \ref{T:1st-moment-01}.
The relative uncertainties become smaller in the set C from the set A:
$\delta\Delta G(x>0.1)/\Delta G(x>0.1)$
= 1.36 $\rightarrow$ 0.73 (0.71 in set B)
in the positive distribution and     
  1.02 $\rightarrow$ 0.54 (0.77 in set B) 
in the node one.

The analysis C has been made with the projected E07-011 data which are
created so as to agree with the set-A spin asymmetries. In actual
experimental measurements, statistical fluctuations exist. Therefore,
the reduction of the gluon uncertainty could be overestimated.
In order to investigate such a possibility, we tried the following
analyses by generating the E07-011 data with two different assumptions:
\begin{itemize}
\item Analysis C$'$: 
           Projected E07-011 data are obtained by using the node-type
           distributions in the set A, and then a global analysis
           is made with positive-type initial distributions.
\item Analysis C$''$:
           Projected E07-011 data are obtained by using the positive-type
           distributions in the set A, and then a global analysis
           is made with node-type initial distributions.
\end{itemize}

\begin{table}[t]
\caption{
The first moment $\Delta G$ and its uncertainty
         $\delta \Delta G$ at $Q^2=1$ GeV$^2$ in the range $0.1<x<1$.
         The notations are the same as the ones in 
         Table \ref{T:1st-moment-01}.
         }
\label{T:1st-moment-01-c}
\vspace{2mm}
\begin{center}
\begin{tabular}{ccccc}
\hline
\ Function \ & \ Set \
             &   $\Delta G (x>0.1)$         \ 
             &   $\delta \Delta G (x>0.1)$  \
             &   $\frac{\delta \Delta G (x>0.1)}{\Delta G (x>0.1)}$ \ \\
\hline
Positive     & C$'$   & 0.49   & 0.47   & 0.96 \\
Node         & C$''$  & 0.74   & 0.52   & 0.70 \\
\hline
\end{tabular}
\end{center}
\end{table}

Analysis results of C$'$ indicate that the polarized PDFs and 
their uncertainties are almost the same as the positive-type
distributions of the analysis C in the valence-quark and antiquark parts.
Determined PDFs of C$''$ are also almost the same as the node-type ones
of the set C. However, the gluon distributions and their uncertainties
are slightly modified although $x$-dependent functional forms are
not changed. The gluon first moments and their uncertainties are
shown in Table \ref{T:1st-moment-01-c}. 
As expected, the uncertainties become larger than the ones for
the analysis C in Table \ref{T:1st-moment-01}:
$ \delta \Delta G (x>0.1) / \Delta G (x>0.1)=0.96$ 
      [0.73 in set C; 0.71 (B); 1.36 (A)] for the positive type;
0.70  [0.54 in set C; 0.77 (B); 1.02 (A)] for the node type.
According to the C$'$  and C$''$ results, the improvement from
the set A is not as large as the one in the set C. However,
it is still comparable with the RHIC improvement in B.

It is important to find that the large uncertainty of $\Delta G$
is significantly reduced by the proposed E07-011 experiment
according to the analysis C (and C$'$, C$''$). The improvement of
$\Delta G$ is especially clear in the integral over the region $x>0.1$.
Therefore, the proposed JLab experiment
is valuable for the determination of $\Delta G$ in addition to
the quark and antiquark moments. If the polarized lepton scattering
data are accurate enough, they can contribute to a better determination
of the gluon spin content.
It is interesting to find that the analysis-C uncertainties are almost
the same as the ones in the analysis B. However, it should be noted that
the uncertainties of the pion fragmentation functions and scale
uncertainties are not taken into account in the error estimate
of the analysis B.

From these comparisons, we found that the JLab measurements of
the the proposed E07-011 can contribute to reducing the uncertainty
of the gluon spin content by about 47\% in the analysis C
(30\% in C$'$ and C$''$).
The RHIC $\pi^0$ measurements also reduce the uncertainty, but
the inclusive lepton scattering data are valuable in the sense
that they are free from the ambiguities of the fragmentation
functions \cite{hkns07}. 

We should, however, mention that future RHIC
experiments will also improve the determination of $\Delta g(x)$
by measurements of pion, jet, and direct-photon production processes.
To be fair with the RHIC-Spin and other high-energy spin projects,
it is desirable that a global analysis should be made together
with other expected data by the time of the future JLab-E07-011 data.
Furthermore, the current estimation would indicate a slightly better
determination of $\Delta g(x)$ from the JLab data because
some fluctuations are expected in actual measurements from the cental
spin asymmetries. However, we believe that our analyses with
the expected errors are good enough in comparing the obtained
relative errors for the gluon spin content from the DIS and
collider data. In this article, we simply showed the role of
accurate DIS data, by taking the JLab-E07-011 data as an example,
in comparison with the current status because all the expected
data are not available for our analysis and also it is not easy
to judge which data will be taken before the JLab-E07-011 data.

Because it is one of our purposes to provide a polarized PDF library,
which has not been provided since 2003, we did not include preliminary
data such as the RHIC run-6 data in our global analyses. In addition,
semi-inclusive data are also not included because we focused our
analysis on the determination of $\Delta g(x)$ and the results
depend much on used fragmentation functions. In future, we expect to
make a more complete analysis including the RHIC run-6 and semi-inclusive
data. 

\section{Summary}
\label{summary}

We have investigated the impacts of $\pi^0$-production data in polarized
$pp$ collisions at RHIC and measurements of $g_1^d$ in the future JLab
experiment E07-011 on the determination of the polarized PDFs, especially
the polarized gluon distribution. Global analyses of the polarized DIS
were done for determining the polarized PDFs and their uncertainties
by using the DIS data on $g_1$, the $\pi^0$ data of the RHIC RUN-5,
and the expected JLab-E07-011 data on $g_1^d$.

The RHIC $\pi^0$ data indicated a possibility of a node-type
distribution for $\Delta g(x)$ which changes sign at $x\sim 0.1$.
Accordingly, we made two types of analyses with positive and node-type
distributions for $\Delta g(x)$. Similar $\chi^2_{min}$ values
were obtained by both analyses although the node-type distributions
have slightly smaller $\chi^2_{min}$. It suggests that it is
difficult to determine a precise gluon distribution from the current data.
However, it is interesting to find in our analyses that the distribution
$\Delta g(x)$ is positive at $x>0.2$ because of scaling violation
in $g_1$ and $\pi^0$-production data.
We found a large uncertainty for $\Delta g(x)$ at small $x$ ($<0.1$).
The uncertainty of the polarized gluon polarization is significantly
reduced by the RHIC $\pi^0$ data. In particular, the $\pi^0$ data
constrain the gluon moment calculated in the large-$x$ region
($x>0.1$), and the reduction of the gluon uncertainty is 24\%$\sim$47\%
according to our analysis. 

There is a clear improvement in the determination of $\Delta g(x)$
by the future measurements of the E07-011 experiment. 
It is partially due to the error correlation between the polarized
antiquark and gluon distributions. However, the major improvement
comes from the fact that the measurements are accurate enough to probe
the polarized gluon distribution in the structure function $g_1$.
Namely, the experimental errors are much smaller than the typical
gluonic NLO term in $g_1$.
The E07-011 data can reduce the uncertainty of the first moment
$\Delta G$ by about 30$-$47\% if only the inclusive DIS data are used
in the analysis. This reduction is comparable to the effect of
the RHIC RUN-5 $\pi^0$ data. However, the JLab E07-011 data could
have an advantage over the hadron-production data because
there is no uncertainty coming from the fragmentation functions
and the choice of the hard scale. A better determination of $\Delta g(x)$
should be also made by future measurements of pion, jet, and direct-photon
production processes at RHIC.

Precise DIS and collider measurements are important for
determining the polarized PDFs, especially in the polarized
gluon distribution. From the analyses B,
we provide a code for calculating two sets of optimum polarized PDFs.
The set-1 is for the positive-type PDFs in the analysis B,
and the set-2 is for the node-type ones also in the analysis B.
The code is provided at our web home page \cite{aac-web}.

\section*{Acknowledgments}
The authors thank N. Saito for useful discussions and comments.
They also thank P. Bosted, X. Jiang, and S. Kuhn for communications
on JLab experiments
and S. Albino, J. Bl\"umlein, H. B\"ottcher, A. Miller, and Y. Miyachi 
for communications on the error analysis. 
They were partially supported by the Grant-in-Aid
for Scientific Research from the Japanese Ministry
of Education, Culture, Sports, Science, and Technology. 


\end{document}